\documentclass[12pt]{article}
\usepackage{epsfig}
\def\be{\begin{equation}}
\def\ee{\end{equation}}
\def\bea{\begin{eqnarray}}
\def\eea{\end{eqnarray}}
\usepackage{graphicx}% Include figure files

\catcode`\@=11
\def\lsim{\mathrel{\mathpalette\@versim<}}
\def\gsim{\mathrel{\mathpalette\@versim>}}
\def\@versim#1#2{\vcenter{\offinterlineskip
\ialign{$\m@th#1\hfil##\hfil$\crcr#2\crcr\sim\crcr } }}
\catcode`\@=12

\catcode`@=12
\begin{document}
\thispagestyle{empty}
\begin{flushright}
UCRHEP-T552\\
June 2015\
\end{flushright}
\vspace{0.6in}
\begin{center}
{\LARGE \bf Transformative A$_4$ Mixing of Neutrinos\\ 
with CP Violation\\}
\vspace{1.2in}
{\bf Ernest Ma\\}
\vspace{0.2in}
{\sl Physics \& Astronomy Department and Graduate Division,\\ 
University of California, Riverside, California 92521, USA\\}
\end{center}
\vspace{1.2in}

\begin{abstract}\
A new theoretical insight into the pattern of neutrino mixing and leptonic 
$CP$ violation is presented.  It leads naturally and uniquely to a specific 
dark sector of three real neutral scalar singlets, with the radiative 
implementation of the inverse seesaw mechanism for neutrino mass.  The new 
simple but crucial enabling idea is that a familiar $A_4$ transformation 
turns any orthogonal $3 \times 3$ matrix into one which predicts $\theta_{23} 
= \pi/4$ and $\delta_{CP} = \pm \pi/2$ for the neutrino mixing matrix, in 
good agreement with present data.
\end{abstract}

\newpage
\baselineskip 24pt

In recent years, many theoretical studies have been made regarding the 
pattern of the $3 \times 3$ neutrino mixing matrix.  In particular, 
the use of non-Abelian discrete symmetries is widespread.  This came about 
from the specific example of $A_4$~\cite{mr01,m02,bmv03}, where it 
was shown for the first time how the three very different charged-lepton 
masses may be incorporated into a symmetry for neutrino mixing, which 
can explain $\sin^2 \theta_{23} = 1/2$.  Subsequently, motivated by empirical 
observation, it was conjectured~\cite{hps02} that the pattern could 
be tribimaximal, with $\sin^2 \theta_{12} = 1/3$ and $\theta_{13} = 0$.  It 
was then shown~\cite{m04} that $A_4$ is indeed suitable for obtaining this 
result.  Since 2005, there have been many papers written regarding this 
possibility.

In 2011~\cite{t2k11} and then more decisively in 2012~\cite{dbay12,reno12}, 
$\theta_{13}$ was measured to be significantly different from zero, thus 
falsifying the tribimaximal ansatz.  The 2014 Particle Data Group (PDG) 
values~\cite{pdg2014} of neutrino parameters are:
\begin{eqnarray}
&& \sin^2 (2 \theta_{12}) = 0.846 \pm 0.021, ~ \Delta m^2_{21} = (7.53 \pm 
0.18) \times 10^{-5}~{\rm eV}^2, \\ 
&& \sin^2 (2 \theta_{23}) = 0.999 \pmatrix{+0.001 \cr -0.018}, ~  
\Delta m^2_{32} = (2.44 \pm 0.06) \times 10^{-3}~{\rm eV}^2 ~({\rm normal}), \\ 
&& \sin^2 (2 \theta_{23}) = 1.000 \pmatrix{+0.000 \cr -0.017}, ~  
\Delta m^2_{32} = (2.52 \pm 0.07) \times 10^{-3}~{\rm eV}^2 ~({\rm inverted}), \\
&& \sin^2 (2 \theta_{13}) = (9.3 \pm 0.8) \times 10^{-2},
\end{eqnarray}
where (normal) refers to the ordering $m_1 < m_2 < m_3$ of neutrino masses, and 
(inverted) refers to $m_3 < m_1 < m_2$.

More recently~\cite{t2k15}, combining reactor data, there appears to be 
a preference for $\delta_{CP} = -\pi/2$ in long-baseline neutrino oscillation 
data.  These new developments are in fact consistent with a special form of 
the Majorana neutrino mass matrix which first appeared in 
2002~\cite{bmv03,m02-1}, i.e.
\begin{equation}
{\cal M}_\nu^{(e, \mu, \tau)} = \pmatrix{A & C & C^* \cr C & D^* & B \cr C^* 
& B & D}, 
\end{equation}
where $A,B$ are real.  This allows $\theta_{13} \neq 0$ and yet $\theta_{23} 
= \pi/4$ is maintained, together with the prediction that $\delta_{CP} = \pm 
\pi/2$.  Subsequently, this pattern was shown~\cite{gl04} to be protected by 
a symmetry, i.e. $e \to e$ and $\mu \leftrightarrow \tau$ exchange with 
$CP$ conjugation.  With the knowledge that $\theta_{13} \neq 0$, this 
extended symmetry is now the subject of many studies, which 
began with generalized $S_4$~\cite{mn12}.

In this paper, I show how $\theta_{23} = \pi/4$ and $\delta_{CP} = \pm 
\pi/2$ may be obtained in a very general way, 
using the familiar unitary $3 \times 3$ transformation
\begin{equation}
U_\omega = {1 \over \sqrt{3}} \pmatrix{1 & 1 & 1 \cr 1 & \omega & \omega^2 \cr 
1 & \omega^2 & \omega},
\end{equation}
where $\omega = \exp(2 \pi i/3) = -1/2 + i \sqrt{3}/2$, which is derivable 
from $A_4$ as shown in Ref.~\cite{mr01}.  The idea is very simple.  
Consider the product of $U_\omega {\cal O}$, where ${\cal O}$ is a real 
orthogonal $3 \times 3$ matrix, i.e.
\begin{equation}
{1 \over \sqrt{3}} \pmatrix{1 & 1 & 1 \cr 1 & \omega & \omega^2 \cr 
1 & \omega^2 & \omega} \pmatrix{o_{11} & o_{12} & o_{13} \cr o_{21} & 
o_{22} & o_{23} \cr o_{31} & o_{32} & o_{33}} = \pmatrix{u_{11} & u_{12} & 
u_{13} \cr u_{21} & u_{22} & u_{23} \cr u_{31} & u_{32} & u_{33}} = U. 
\end{equation}
It is clear that $u_{2i}^* = u_{3i}$ for $i = 1,2,3$.  Comparing this with 
the PDG convention of the neutrino mixing matrix, i.e.
\begin{equation}
U = \pmatrix{c_{12} c_{13} & s_{12} c_{13} & s_{13} e^{-i \delta} \cr 
-s_{12} c_{23} -c_{12} s_{23} s_{13} e^{i \delta} & c_{12} c_{23} -s_{12} 
s_{23} s_{13} e^{i \delta} & s_{23} c_{13} \cr 
s_{12} s_{23} -c_{12} c_{23} s_{13} e^{i \delta} & -c_{12} s_{23} -s_{12} 
c_{23} s_{13} e^{i \delta} & c_{23} c_{13}}, 
\end{equation}
it is obvious that after rotating the phases of the third column and the 
second and third rows, the two matrices are identical if and only if 
$s_{23} = c_{23}$ and $\cos \delta = 0$, i.e. $\theta_{23} = \pi/4$ and 
$\delta_{CP} = \pm \pi/2$.  This was first pointed out in 
Refs.~\cite{fmty00,mty01}.

To obtain this result, the necessary condition is that the $3 \times 3$ 
Majorana neutrino mass matrix ${\cal M}_\nu$ must be diagonalized by an 
orthogonal matrix in the $A_4$ basis.  Obviously it will be so if 
${\cal M}_\nu$ is purely real.  In that case, in the 
$(e,\mu,\tau)$ basis, it is given by
\begin{equation}
{\cal M}_\nu^{(e,\mu,\tau)} = U_\omega \pmatrix{a & c & e \cr c & d & b 
\cr e & b & f} U^T_\omega= \pmatrix{A & C & C^* \cr C & D^* & B \cr 
C^* & B & D},
\end{equation}
where
\begin{eqnarray}
A &=& (a + 2b + 2c + d + 2e + f)/3, \\ 
B &=& (a - b - c + d - e + f)/3, \\ 
C &=& (a - b - \omega^2 c + \omega d - \omega e + \omega^2 f)/3, \\ 
D &=& (a + 2b + 2 \omega^2 c + \omega d + 2 \omega e + \omega^2 f)/3.
\end{eqnarray}
In other words, the form of Eq.~(5) is automatically obtained, as expected.

In the context of $A_4$, efforts prior to 2011 were concentrated on how to 
achieve $c = e = 0$ and $d = f$ for tribimaximal mixing without a necessarily 
real ${\cal M}_\nu$, i.e. a residual 
$Z_2$ symmetry in the neutrino sector which coexists with the residual 
$Z_3$ symmetry implied by $U_\omega$ in the charged-lepton sector.  This 
clash or misalignment of residual symmetries is the origin of a basic 
theoretical problem which has no simple solution.  In hindsight, it 
is a powerful argument against the naive expectation of an exact tribimaximal 
form of the neutrino mixing matrix.  Here $A_4$ serves simply as a 
link for a (real) neutrino mass matrix without any symmetry to the 
charged-lepton sector.  The new remarkable result is that the nature of this 
link, i.e. $U_\omega$, leads to two verifiable specific predictions, i.e. 
$\theta_{23} = \pi/4$ and $\delta_{CP} = \pm \pi/2$, which agree well with 
present data.  In Ref.~\cite{h15}, $c=e=0$ is again assumed but $d$ and $f$ 
are not set equal.  In this way $\theta_{13} \neq 0$ is obtained and the 
further assumption (but without further justification) that $a, b, d, f$ 
are real leads to $\theta_{23} = \pi/4$ and $\delta_{CP} = \pm \pi/2$.

Here I answer the new important question of how an arbitrary complex neutrino 
mass matrix can be guaranteed to be purely real, without imposing explicit 
$CP$ conservation.  The key is of course the origin of ${\cal O}$ which 
obviously would be the result of diagonalizing a real $3 \times 3$ mass 
matrix.  The only guaranteed such mass matrix is that of three real scalars.  
Hence the quest for ${\cal O}$ leads inexorably to a mechanism by which 
neutrino masses come from three real scalars.  This is the significance of 
Eq.~(7).  In the following, I will show that it may be achieved 
naturally together with the appearance of $U_\omega$ in a radiative 
implementation~\cite{m06,m14} of neutrino and charged-lepton masses through 
dark matter (scotogenic), using {\it only} the one Higgs doublet of the 
standard model (SM), as suggested by the observation~\cite{atlas12,cms12} 
of the 125 GeV particle at the Large Hadron Collider (LHC).

Under $A_4$, let the three families of leptons transform as
\begin{equation}
(\nu_i,l_i)_L \sim \underline{3}, ~~~ l_{iR} \sim \underline{1}, 
\underline{1}', \underline{1}''.
\end{equation}
Add the following new particles, all assumed odd under an exactly conserved 
discrete $Z_2$ (dark) symmetry, whereas all SM particles are even:
\begin{equation}
(E^0,E^-)_{L,R} \sim \underline{1}, ~~~ N_{L,R} \sim \underline{1}, ~~~ 
s_i \sim \underline{3}, 
\end{equation}
where $(E^0,E^-)$ is a fermion doublet, $N$ a neutral fermion singlet, and 
$s_{1,2,3}$ are {\it real} neutral scalar singlets.  Together with the one 
Higgs doublet $(\phi^+,\phi^0)$ of the SM, one-loop radiative inverse 
seesaw neutrino masses are generated~\cite{fmp14,mnp15} as shown in Fig.~1.
\begin{figure}[htb]
\vspace*{-3cm}
\hspace*{-3cm}
\includegraphics[scale=1.0]{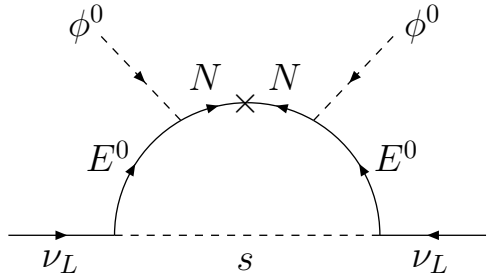}
\vspace*{-21.5cm}
\caption{One-loop generation of inverse seesaw neutrino mass.}
\end{figure}

These new terms in the Lagrangian are given by
\begin{eqnarray}
{\cal L}' &=& -m_N \bar{N} N - m_E (\bar{E}^0 E^0 + \bar{E}^- E^-) - 
{1 \over 2} m_L N_L N_L - {1 \over 2} m_R N_R N_R + {1 \over 2} (m_s^2)_{ij} 
s_i s_j \nonumber \\ 
&+& f_D \bar{N}_L (E_R^0 \phi^0 - E_R^- \phi^+) + f_F \bar{N}_R (E_L^0 
\phi^0 - E_L^- \phi^+) + f s_i (\bar{E}^0_R \nu_{iL} + \bar{E}^-_R l_{iL}) 
+ H.c.
\end{eqnarray}
The mass matrix linking $(\bar{N}_L, \bar{E}^0_L)$ to $(N_R, E^0_R)$ 
is then
\begin{equation}
{\cal M}_{N,E} = \pmatrix{m_N & m_D \cr m_F & m_E},
\end{equation}
where $m_D = f_D \langle \phi^0 \rangle$, and $m_F = f_F \langle \phi^0 
\rangle$.  As a result, $N$ and $E^0$ mix to form two Dirac fermions 
of masses $m_{1,2}$, with mixing angles
\begin{eqnarray}
m_D m_E + m_F m_N &=& \sin \theta_L \cos \theta_L (m_1^2 - m_2^2), \\
m_D m_N + m_F m_E &=& \sin \theta_R \cos \theta_R (m_1^2 - m_2^2).
\end{eqnarray}
To connect the loop, Majorana mass terms $m_L$ and $m_R$ are necessary. 
Since both $E$ and $N$ may be defined to carry lepton number, these terms 
violate lepton number softly and may be naturally small, thus realizing 
the mechanism of inverse seesaw~\cite{ww83,mv86,m87}.  The one-loop 
Majorana neutrino mass is given by 
\begin{eqnarray}
m_\nu &=& f^2 m_R \sin^2 \theta_R \cos^2 \theta_R (m_1^2 - m_2^2)^2 
\int {d^4 k \over (2 \pi)^4} {k^2 \over (k^2 - m_s^2)} {1 \over 
(k^2 - m_1^2)^2} {1 \over (k^2 - m_2^2)^2} \nonumber \\ 
&+& f^2 m_L m_1^2 \sin^2 \theta_R \cos^2 \theta_L \int {d^4 k \over (2 \pi)^4} 
{1 \over (k^2 - m_s^2)}{1 \over (k^2 - m_1^2)^2} \\ 
&+& f^2 m_L m_2^2 \sin^2 \theta_L \cos^2 \theta_R \int {d^4 k \over (2 \pi)^4}
{1 \over (k^2 - m_s^2)}{1 \over (k^2 - m_2^2)^2} \nonumber \\ 
&-& 2 f^2 m_L m_1 m_2 \sin \theta_L \sin \theta_R \cos \theta_L \cos \theta_R 
\int {d^4 k \over (2 \pi)^4}{1 \over (k^2 - m_s^2)}{1 \over (k^2 - m_1^2)}
{1 \over (k^2 - m_2^2)}. \nonumber  
\end{eqnarray}
This formula holds for $s$ as a mass eigenstate.  If $A_4$ is unbroken, then 
$s_{1,2,3}$ all have the same mass and ${\cal M}_\nu$ is proportional to the 
identity matrix.  However, if $A_4$ is softly broken by the necessarily 
real $s_i s_j$ mass terms, then the neutrino mass matrix is given by
\begin{equation}
{\cal M}_\nu = {\cal O} \pmatrix{m_{\nu1} & 0 & 0 \cr 0 & m_{\nu2} & 0 \cr 
0 & 0 & m_{\nu3}} {\cal O}^T,
\end{equation}
where ${\cal O}$ is an orthogonal matrix. 
Now each $m_{\nu i}$ may be complex because $f$, $m_L$, $m_R$ may be 
complex in Eq.~(20), but a common unphysical phase, say for $\nu_1$, 
may be rotated away, leaving just two relative Majorana phases for 
$\nu_2$ and $\nu_3$, owing to the relative phase between $m_L$ and 
$m_R$ with different $s_{1,2,3}$ masses in Eq.~(20).  Hence ${\cal M}_\nu$ 
is diagonalized by ${\cal O}$, which is all that is required to obtain 
$\theta_{23} = \pi/4$ and $\delta_{CP} = \pm \pi/2$, once $U_\omega$ is 
applied.  This shows that the neutrino mass matrix does not have to be real.  
It only has to be diagonalized by an orthogonal matrix.

To derive $U_\omega$, the simplest way is to copy Ref.~\cite{mr01} and 
add three Higgs doublets $\Phi_i \sim \underline{3}$.  This leads to 
the charged-lepton mass matrix
\begin{eqnarray}
{\cal M}_l &=& \pmatrix{f_e v_1^* & f_\mu v_1^* & f_\tau v_1^* \cr 
f_e v_2^* & f_\mu \omega^2 v_2^* & f_\tau \omega v_2^* \cr 
f_e v_3^* & f_\mu \omega v_3^* & f_\tau \omega^2 v_3^*} \nonumber \\ 
&=& \pmatrix{v_1^* & 0 & 0 \cr 0 & v_2^* & 0 \cr 0 & 0 & v_3^*} 
\pmatrix{1 & 1 & 1 \cr 1 & \omega^2 & \omega \cr 1 & \omega & \omega^2} 
\pmatrix{f_e & 0 & 0 \cr 0 & f_\mu & 0 \cr 0 & 0 & f_\tau}.
\end{eqnarray}
For $v_1 = v_2 = v_3$, a residual $Z_3$ symmetry exists with $m_e = \sqrt{3} 
f_e v$, etc. and $U_\omega$ becomes the transformation linking 
${\cal M}_l$ to ${\cal M}_\nu$.  However, this scenario requires four 
Higgs doublets.  It is thus somewhat problematic in the face of present 
data regarding the observed~\cite{atlas12,cms12} 125 GeV particle, 
which is entirely consistent with being the one Higgs boson $h$ of the SM.

To obtain charged-lepton masses in the context of $A_4$ with just the SM 
Higgs doublet, the general radiative framework of Ref.~\cite{m14} is adopted. 
The specific scenario here requires the addition of two sets of charged 
scalars odd under dark $Z_2$:
\begin{equation}
x^-_i \sim \underline{3}, ~~~ y^-_i \sim \underline{1}, \underline{1}', 
\underline{1}''. 
\end{equation}
The one-loop diagram is given in Fig.~2.
\begin{figure}[htb]
\vspace*{-3cm}
\hspace*{-3cm}
\includegraphics[scale=1.0]{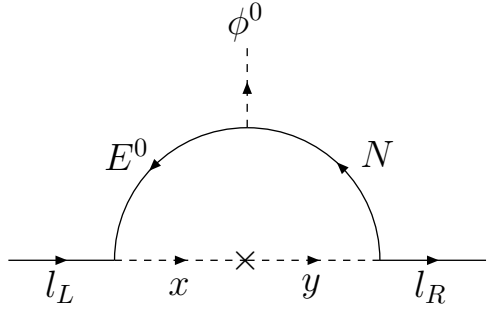}
\vspace*{-21.5cm}
\caption{One-loop generation of charged-lepton mass.}
\end{figure}
To connect $x$ with $y$, $A_4$ must be broken, either softly so that 
the link is again $U_\omega$ to obtain the desired residual $Z_3$ symmetry, 
or spontaneously using three singlet scalar 
fields $\chi_i \sim \underline{3}$ with equal vacuum expectation values. 
In this way, the three Higgs doublets of the original $A_4$ model are 
replaced in a renormalizable theory for obtaining charged-lepton masses. 
Note that the latter may be considered as the ultraviolet completion of the 
common practice of using the nonrenormalizable dimension-five term 
$\bar{l}_L l_R \bar{\phi}^0 \chi$ for such a purpose.

As a result, the charged-lepton mass matrix is given by
\begin{equation}
{\cal M}_l = U^\dagger_\omega \pmatrix{m_e & 0 & 0 \cr 0 & m_\mu & 0 \cr 
0 & 0 & m_\tau},
\end{equation}
with
\begin{equation}
m_e = f' f_e \mu_e u \int {d^4 k \over (2 \pi)^4} {1 \over (k^2 - m_{1e}^2)
(k^2 - m_{2e}^2)} \left[ {m_1 \cos \theta_R \sin \theta_L \over k^2 - m_1^2} 
- {m_2 \cos \theta_L \sin \theta_R \over k^2 - m_2^2} \right],
\end{equation}
where $f'$ is the $E^0_L l_L x^*$ Yukawa coupling, $f_e$ is the $N_R e_R y^*_1$ 
Yukawa coupling, $\mu_e$ is the scalar trilinear $x y_1^* \chi$ coupling, 
$u$ is the vacuum expectation value of $\chi$, and $m_{1e,2e}$ are the mass 
eigenvalues of the $2 \times 2$ mass-squared matrix
\begin{equation}
{\cal M}^2_{xy_1} = \pmatrix{m_x^2 & \mu_e u \cr \mu_e u & m_{y_1}^2},
\end{equation}
with $\mu_e u = \sin \theta_e \cos \theta_e (m^2_{1e} - m^2_{2e})$, 
and similarly for $m_\mu$ 
and $m_\tau$.  One immediate consequence of a radiative charged-lepton mass 
is that the Higgs Yukawa coupling $h \bar{l} l$ is no longer exactly $m_l/v$ 
as in the SM.  Its deviation is not suppressed by the usual one-loop 
factor of $16 \pi^2$ and may be large enough to be observable~\cite{fm14}. 

There is a one-to-one correlation of the neutrino mass 
eigenstates to the $s_{1,2,3}$ mass eigenstates, the lightest of which is 
dark matter~\cite{sz85,cskw13}.  It is also clear from Eq.~(20) that all 
three neutrino masses are expected to be of the same order of magnitude, 
and their mass-squared differences are related to the scalar mass differences. 
The most recent cosmological data~\cite{planck15} imply
\begin{equation}
\sum m_\nu < 0.23~{\rm eV}.
\end{equation}
This would mean that the effective neutrino mass $m_{ee}$ in neutrinoless 
double beta decay is bounded below 0.07 eV for normal ordering and 0.08 eV 
for inverted ordering.

Due to the presence of the $A_4$ symmetry, the dark matter parity of this 
model is also derivable from lepton parity~\cite{m15}.  Under lepton 
parity, let the new particles $(E^0,E^-), N$ be even and $s,x,y$ be odd, 
then the same Lagrangian is obtained.  As a result, dark parity is simply 
given by $(-1)^{L+2j}$, which is odd for all the new particles and even 
for all the SM particles.  Note that the tree-level Yukawa 
coupling $\bar{l}_L l_R \phi^0$ would be allowed by lepton parity alone, 
but is forbidden here because of the $A_4$ symmetry.  The lightest $s$ is 
dark matter.  If its relic density as well as direct-detection cross section 
are determined only by the $\lambda s^2 (\Phi^\dagger \Phi)$ interaction,  
then its allowed parameter space is limited to a small region just below 
$m_h/2$~\cite{fpu15}.  In the above model, if the interaction of $s$ with 
$\chi$ is also taken into account, it adds to the $ss$ annihilation cross 
section but not to the elastic scattering of $s$ off nuclei.  The former 
may then satisfy the relic abundance requirement and yet the latter will 
evade the direct-detection constraint.  Details will be presented elsewhere.

In conclusion, it has been pointed out that the phenomenologically 
successful values of $\theta_{23} = \pi/4$ and $\delta_{CP} = \pm \pi/2$ 
for the neutrino mixing matrix is derivable from the familiar $A_4$ 
transformation of Eq.~(6) if it is multiplied by an orthogonal matrix. 
This leads to the specific notion that a desirable neutrino mass matrix 
should come from three real scalars in the context of $A_4$. 
To obtain the latter naturally, a specific 
scotogenic one-loop radiative model of neutrino and charged-lepton masses is 
proposed, where the particles appearing in the loop have odd dark matter 
parity.  These predicted new particles should have masses at the scale of  
weakly interacting dark matter, i.e.  1 TeV or less, and be potentially 
observable at the LHC, which has just resumed operation at CERN.

\medskip
This work is supported in part 
by the U.~S.~Department of Energy under Grant No.~DE-SC0008541.

\bibliographystyle{unsrt}

\end{document}